\documentclass[12pt]{article}
\usepackage[margin=1in]{geometry}
\usepackage{setspace}

\usepackage{float}
\usepackage{amsfonts}

\usepackage{amssymb}
\usepackage{latexsym}
\usepackage{graphicx}
\usepackage[english]{babel}
\usepackage[font={small}]{caption}

\usepackage{amsfonts}
\usepackage{latexsym}
\usepackage{graphicx}
\usepackage[english]{babel}
\usepackage{tikz}

\begin{document}
\input epsf

\def\p{\partial}
\def\h{{1\over 2}}
\def\be{\begin{equation}}
\def\bea{\begin{eqnarray}}
\def\ee{\end{equation}}
\def\eea{\end{eqnarray}}
\def\d{\partial}
\def\la{\lambda}
\def\eps{\epsilon}
\def\bb{\bigskip}
\def\mm{\medskip}
\newcommand{\dm}{\begin{displaymath}}
\newcommand{\edm}{\end{displaymath}}
\renewcommand{\b}{\tilde{B}}
\newcommand{\gm}{\Gamma}
\newcommand{\ac}[2]{\ensuremath{\{ #1, #2 \}}}
\renewcommand{\ell}{l}
\newcommand{\z}{\ell}
\newcommand{\newsection}[1]{\section{#1} \setcounter{equation}{0}}
\def\bb{$\bullet$}
\def\Qbar{{\bar Q}_1}
\def\QPbar{{\bar Q}_p}

\def\q{\quad}

\def\bn{B_\circ}

\let\a=\alpha \let\b=\beta \let\g=\gamma \let\d=\delta \let\e=\epsilon
\let\c=\chi \let\th=\theta  \let\k=\kappa
\let\l=\lambda \let\m=\mu \let\n=\nu \let\x=\xi \let\r=\rho
\let\s=\sigma \let\t=\tau
\let\vp=\varphi \let\vep=\varepsilon
\let\w=\omega      \let\G=\Gamma \let\D=\Delta \let\Th=\Theta
                     \let\P=\Pi \let\S=\Sigma

\def\h{{1\over 2}}
\def\t{\tilde}
\def\r{\rightarrow}
\def\nn{\nonumber\\}
\let\bm=\bibitem
\def\Kt{{\tilde K}}
\def\b{\bigskip}
\def\m{\medskip}

\let\p=\partial

\newcommand\blfootnote[1]{%
  \begingroup
  \renewcommand\thefootnote{}\footnote{#1}%
  \addtocounter{footnote}{-1}%
  \endgroup
}

\newcounter{daggerfootnote}
\newcommand*{\daggerfootnote}[1]{%
    \setcounter{daggerfootnote}{\value{footnote}}%
    \renewcommand*{\thefootnote}{\fnsymbol{footnote}}%
    \footnote[2]{#1}%
    \setcounter{footnote}{\value{daggerfootnote}}%
    \renewcommand*{\thefootnote}{\arabic{footnote}}%
    }

\begin{flushright}
\end{flushright}
\vspace{20mm}
\begin{center}
{\LARGE The secret structure of the gravitational vacuum\daggerfootnote{Essay awarded second prize in the Gravity Research Foundation 2024 Awards for Essays on Gravitation.}
 }

\vspace{18mm}
{\bf Samir D. Mathur$^{1}$ }

\blfootnote{$^{1}$ email: mathur.16@osu.edu }

\vspace{4mm}

\b

Department of Physics

 The Ohio State University
 
Columbus,
OH 43210, USA

\b

\vspace{4mm}
\end{center}
\vspace{10mm}
\thispagestyle{empty}
\begin{abstract}

We argue that the vacuum of quantum gravity must contain a hierarchical structure of correlations spanning all length scales. These correlated domains (called `vecros') correspond to virtual fluctuations of black hole microstates. Larger fluctuations are suppressed by their larger action, but this suppression is offset by a correspondingly larger phase space of possible configurations. We give an explicit lattice model of these vecro fluctuations, noting how their distribution changes as the gravitational pull of a star becomes stronger. At the threshold of formation of a closed trapped surface, these virtual fluctuations transition into on-shell black hole microstates (fuzzballs). Fuzzballs radiate from their surface like  normal bodies, resolving the information paradox. We also argue that any model without vecro-type extended vacuum correlations cannot resolve the  paradox.

\end{abstract}
\vskip 1.0 true in

\newpage
\setcounter{page}{1}

\doublespace


Black holes have laid  a trap for us. If we use our conventional understanding of general relativity and quantum theory, then we are led to a contradiction. To escape this  trap we must perform a conjuring trick. In this essay we spell out what this  trick must be. Like all conjuring tricks, once  explained, it will hopefully appear simple and natural. 

\m

{\bf {The trap:}}

\m

Consider the gravitational collapse of a star. The curvatures are low everywhere at the point where the star falls through its horizon radius $r_h=2GM$, so we are forced to agree that the star shrinks uneventfully  through its horizon. New physics can certainly arise at the singularity where the star shrinks to planck size.  But inside the horizon,  light cones  `point inwards'. Thus if we accept that causality holds to leading order in any region with low curvature, then any new physics at the singularity cannot affect dynamics at  horizon, at least to leading order. Furthermore, we can study the evolution using a set of `good slices', which cover the horizon and exterior upto almost the endpoint of evaporation, but which do not approach the singularity and thus have low curvature everywhere. We are then forced to accept that semiclassical dynamics is valid throughout the evolution along these `good slices'.

 Now the trap has closed. Hawking \cite{hawking} found that the quantum  vacuum around the horizon is unstable to the creation of entangled particle pairs $(b_i, c_i)$, whose state  can be schematically modeled as
\be
|\psi\rangle_{pair}={1\over \sqrt{2}}\left ( |0\rangle_b|0\rangle_c+ |1\rangle_b|1\rangle_c\right )
\label{one}
\ee
The $b_i$ escape to infinity as `Hawking radiation', leading to a monotonically growing entanglement of the radiation with the remaining hole. This creates a   a sharp conflict at the endpoint of evaporation.   Hawking's argument was made into a rigorous result by the small corrections theorem \cite{cern}: Any small correction to semiclassical dynamics will not remove the troublesome entanglement; we need an order {\it unity} correction. 
  
  The essential strength of the trap is that curvatures are low all through the evolution along the good slices. It is generally agreed that semiclassical physics must fail when curvatures reach planck scale, but to escape the trap, {\it we need a second mode of failure of semiclassical physics, one where curvatures are low everywhere}. What can give this second mode of failure?
  
\begin{figure}[h]
\begin{center}
\includegraphics[scale=.55]{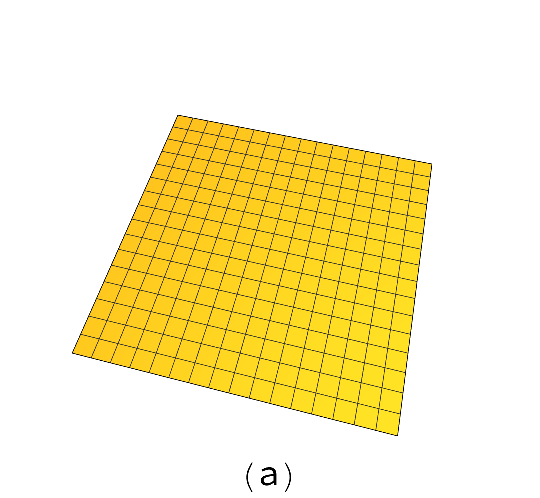}\hskip5em\includegraphics[scale=.55]{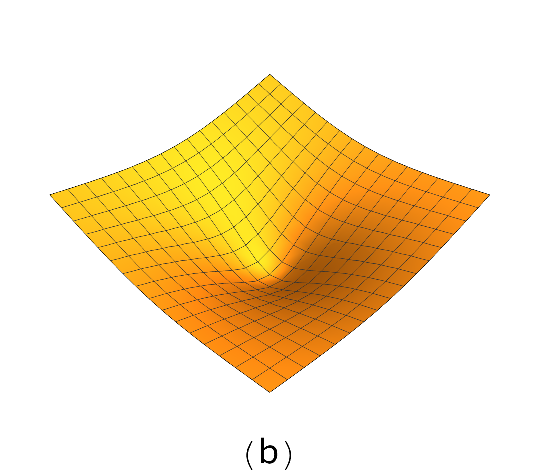}
\end{center}
\caption{(a) Traditional models of quantum gravity assume a complicated structure at the planck scale, but standard low energy effective field theory at larger length scales in any region of gently curved spacetime. (b) As space stretches, we add new points to the grid, maintaining a planck scale lattice.}
\label{f1}
\end{figure}

 To see what will {\it not} work, let us recall the traditional picture of quantum gravity in gently curved space.  Fig.\ref{f1}(a) depicts a region of spacetime with a planck scale grid. We have quantum bits at each lattice site, interacting with the bits on neighboring sites. When spacetime stretches as in fig.\ref{f1}(b), new lattice sites get added to maintain a planck scale grid. The details of the bits and their interaction do not matter; any model in this universality class describes the same low energy semiclassical dynamics. This semiclassical dynamics fails only when curvatures reach planck scale, and so it does not fail along the good slicing of the black hole.   
 
\m

{\bf The role of  vacuum fluctuations:}

\m

To see how to  escape the trap, we first recall a somewhat similar conflict created by Bekenstein's work of 1972. Bekenstein had argued that black holes have an entropy $S_{bek}\sim A/G$ \cite{bek}. Thermodynamics then implies that the hole must radiate at a temperature determined by $TdS_{bek}=dE$. But  light cones  `point inwards' inside the horizon, so nothing should emerge from the region $r<2GM$ to the outside. It seems that the hole cannot radiate, creating a conflict with thermodynamics.

As Hawking showed, once quantum field theory effects are taken into account, black holes do radiate in beautiful agreement with thermodynamics. But we can ask: How do the Hawking quanta  magically appear from empty space?   

The answer to this is of course well known. The vacuum is not empty to begin with; it is full of virtual particle-antiparticle pairs. The collapse of a star distorts  spacetime, and the new geometry has a vacuum with a  different set of virtual  pairs. The difference between the old vacuum and the new one shows up as real (i..e not virtual) pairs; these are the quanta in (\ref{one})  giving Hawking radiation. If we had failed to recognize the existence of these virtual fluctuations in the vacuum wavefunctional, then we would miss the phenomenon of Hawking radiation, and fail to get consistency with thermodynamics.

This history is useful, because in this essay we will argue the following.
{\it We can escape the trap of the information paradox  only if we assume that the gravitational vacuum has a much richer class of vacuum fluctuations than the particle-antiparticle pairs mentioned above.} We will give an explicit lattice model for these fluctuations, and show how they destroy semiclassical evolution in situations where a closed trapped surface forms.  





To understand the full set of fluctuations of the quantum gravity vacuum,  we must first understand the states of the black hole that account for its entropy
 \be
 S_{bek}={A\over 4G}
 \label{three}
 \ee
 In string theory, several classes of  microstates of black holes have been explicitly constructed, and in each case it is found that they have the structure of a {\it fuzzball}, which is like a normal body with no horizon \cite{fuzzballs}. The fuzzball conjecture states that all microstates will be of this form, with the surface of the generic fuzzball expected to be order planck distance outside the horizon radius \cite{ghm}.

\begin{figure}
\begin{center}
\includegraphics[scale=.40]{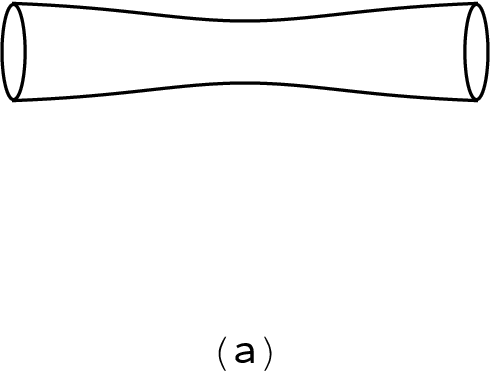}\hskip4em\includegraphics[scale=.40]{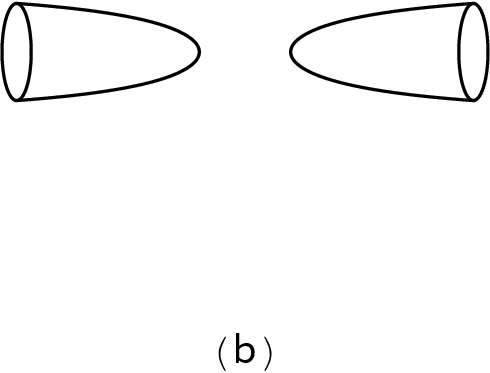}\hskip4em \includegraphics[scale=.40]{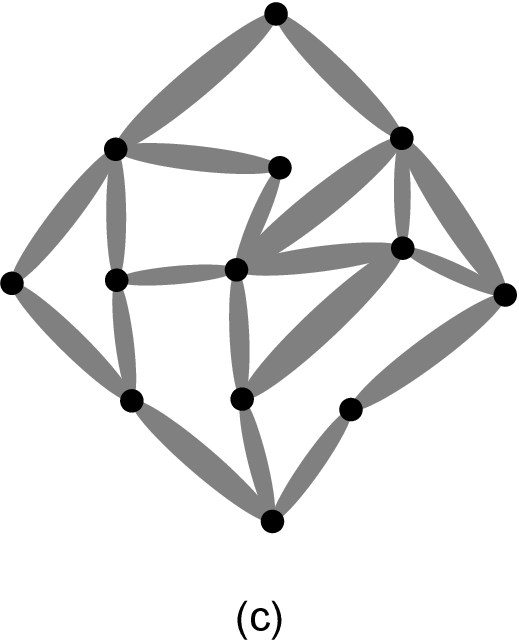}
\end{center}
\caption{(a) Small deformations of compact directions give scalar fields. (b) Larger deformations create topological soliton like Kaluza-Klein monopoles. (c) Black hole microstates are `fuzzballls': complicated bound sets of such solitonic objects.}
\label{f2}
\end{figure}

To understand the structure of fuzzball states, first recall how usual scalar fields arise in string theory. 3+1 dimensional gravity is obtained when we compactify 6 directions to small circles. By the usual picture of Kaluza-Klein reduction, a scalar quantum in the 3+1  theory is obtained as a small deformation of the radius of a compact direction, depicted in fig.\ref{f2}(a). But we can also have a large deformation, depicted in fig.\ref{f2}(b), which changes the local topology of space: The full manifold is no longer a product of 3+1 spacetime with a compact manifold ${\cal M}_6$. In all cases where black hole microstates have been constructed, they are  non-product manifolds of this kind: there is  a complicated structure of KK-monpoles and antimonopoles, with gauge-field fluxes on the topological spheres connecting the monopoles centers.  In \cite{gibbonswarner, prevent} it was shown how such non-product spacetimes invalidate the usual no-hair arguments and allow star-like structures to replace the traditional hole. The fuzzball are `extended' objects which  are very `compression-resistant' due to their planck scale microstructure.

If fuzzballs like fig.\ref{f2}(c) exist as real objects for any mass $M$, then the vacuum must contain virtual fluctuations of these objects. The probability of such a fluctuation is estimated by $P=\left |A \right |^2$, with $A\sim Exp[-S_{grav}]$; here $S_{grav}$ is the gravitational action to create the configuration. Setting all length scales as order $\sim GM$, we find that
\be
S_{grav}\sim {1\over G}\int {\mathcal R}\sqrt{-g} \, d^4 x \sim  GM^2\sim \left ( {M\over m_p}\right)^2
\ee
As expected, $P\ll 1$ for $M\gg m_p$. But this smallness is offset by the very large {\it degeneracy} of fuzzball states of mass $M$ \cite{tunnel}
\be
{\mathcal N} \sim e^{S_{bek}}\sim e^{A\over 4G}\sim e^{4\pi GM^2}= e^{4\pi \left ( {M\over m_p}\right)^2}
\label{tenq}
\ee
Thus we can have
\be
P{\mathcal N}\sim 1
\label{ten}
\ee
{\it Thus the virtual fluctuations of  black hole microstates form an important component of the gravitational vacuum for all masses $0<M<\infty$.} We call these fluctuations {\it vecros}: ``virtual extended compression-resistant objects'' \cite{vecro}.  (Some  applications of the vecro idea are discussed in \cite{elastic,bran}.)

\m

{\bf The conjuring trick we need:}

\m

To escape the trap, we must do the following seemingly impossible things: 

(1)  In the traditional picture of fig.\ref{f1}, {\it nothing} new happens as a star approaches the black hole threshold. In particular, local physics cannot determine if a horizon is about to form. But we {\it must}  find something that becomes important at the black hole threshold!

(2) Since semiclassical physics is expected to be good until this threshold, whatever new effect we find for (1) must be an allowed feature of semiclassical dynamics.

(3) As we cross the black hole threshold, the quantum state should somehow transition to a  fuzzball state of the kind depicted in fig.\ref{f2}(c).

The way our model  will escape the trap is as follows:

(1') The distribution of virtual fluctuations of black hole microstates (vecros) will change as we approach the black hole threshold, peaking at vecros with radius $r\approx 2GM$ at the threshold. The extended nature of the vecro fluctuations allows them to respond to the creation of a closed trapped surface: inside such a surface, the inward pointing structure of light cones squeezes the vecros to make them more compact. 

(2') This change in the vecro distribution alters the vacuum of low energy modes from the Unruh state to the Boulware state. Since the latter is also an allowed state of the semiclassical theory, we satisfy the requirement (2).

(3') As we cross the black hole threshold, the virtual vecro configurations transition to on-shell fuzzballs. Local energy balance is maintained because the Boulware state has negative Casimir energy, while the fuzzball structure has positive energy.


\m

{\bf The model:}

\m

We now  give an explicit lattice model to capture the intuitive argument in (\ref{ten}). 

\begin{figure}
\begin{center}
\includegraphics[scale=.35]{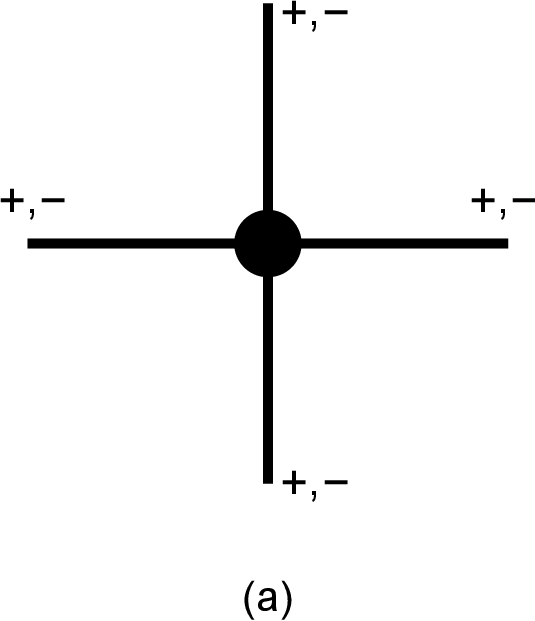}\hskip5em\includegraphics[scale=.35]{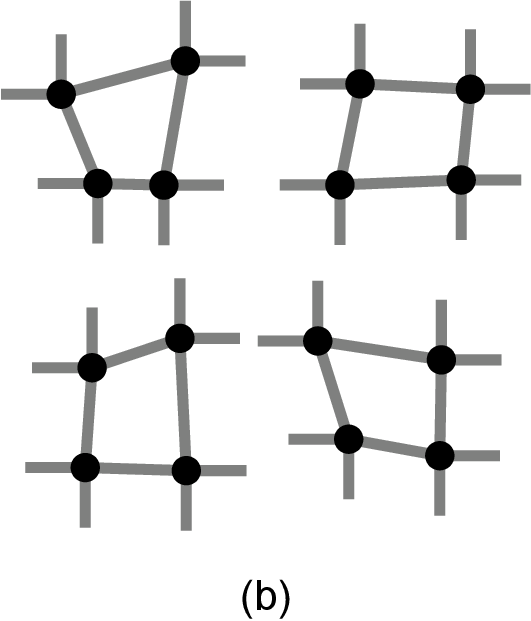}\hskip5em\includegraphics[scale=.35]{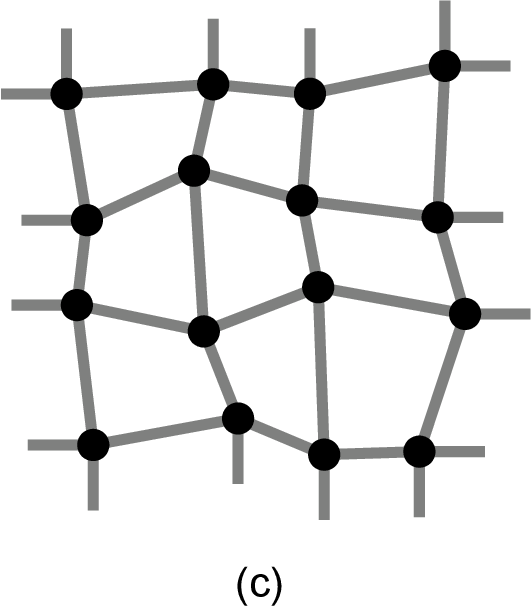}
\end{center}
\caption{(a) A virtual  topological excitation of the type in fig.\ref{f2}(b) is represented by a dot. (b) Such virtual excitations (vecros) can link together to make extended structures. (c) All sizes of vecros are relevant in the vacuum wavefunctional.}
\label{f3}
\end{figure}

(i) We still have the mesh of fig.\ref{f1} describing the fabric of spacetime, but in addition we can have local fluctuations at the lattice sites that are like the bubbles of fig.\ref{f2}(b). Since such bubbles can link together (fig.\ref{f2}(c)), we depict such a bubble in fig.\ref{f3}(a) by a dot with $4$ bonds sticking out from it. These bonds each carry a spin index $\pm$. There is a large phase space of virtual bubble fluctuations of this type, since the bonds on each bubble can have several choices of spins at their ends. But note that there is no entropy associated to such fluctuations; the different spin configurations have a fixed relative amplitude, giving the overall vacuum state $|\Psi_0\rangle$. This is similar to the usual fluctuations of particle-antiparticle pairs in the vacuum, which also do not imply any entropy for the vacuum; the pairs come with fixed relative amplitudes to yield the  unique minimize energy vacuum state $|\Psi_0\rangle$.

\begin{figure}
\begin{center}
\includegraphics[scale=.55]{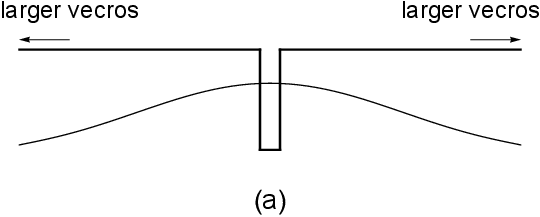}\hskip1em\includegraphics[scale=.55]{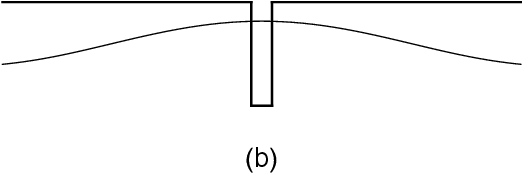}\hskip1em\includegraphics[scale=.55]{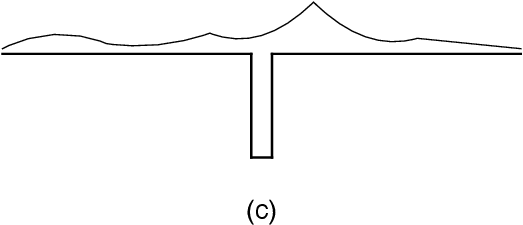}
\end{center}
\caption{(a) The vecro distribution function in the vacuum. (b) The distribution shifts towards larger vecros under the gravitational attraction of a star. (c) Above the black hole threshold, the `under the barrier' virtual wavefunction transitions to `over the barrier' fuzzball states.}
\label{f4}
\end{figure}

(ii) If the bubble fluctuations occur on adjacent lattice sites, then their bonds can link to make a larger structure, as in fig.\ref{f3}(b). Such an extended fluctuation is a model of our `vecro'. Since there are fewer open bonds when bonds link up, the phase space in fig.\ref{f3}(b) is smaller than the phase space where all bubbles were of the type in fig.\ref{f3}(a). But the energy of the configuration is {\it lowered} by this joining of bonds, making the larger vecro fluctuations as relevant as the smaller ones in the vacuum wavefunctional $|\Psi_0\rangle$. Similarly, we have a part of the vacuum wavefunctional with even larger correlated structures  (fig.\ref{f3}(c)). This  hierarchical clustering of virtual `extended' structures models the intuition of eq.(\ref{ten}) that we have virtual fluctuations of extended objects at all scales in the vacuum wavefunctional. Each linked cluster in fig.\ref{f3} is a `vecro'.

(iii) Note that a quantum moving through spacetime does not `bump into' or `scatter off'  virtual fluctuations, as long as the vacuum has approximate local translational invariance. Such fluctuations deform appropriately when a quantum reaches their location and relax to their original form when the quantum has passed through. Our string theory spacetime had 6 compact circles, and scalar fields in the spacetime were described by small fluctuations in the radii $R_i(x), i=1, \dots 6$ of these circles (fig.\ref{f2}(a)). The vecros configurations depend on the $R_i$, and deform accordingly on passage of a quantum. But the only low energy degrees of freedom that can be excited at this stage are the Kaluza-Klein fields arising from the compactification, since the vecro distribution function in the present situation is determined by these low energy fields and is not an independent variable.

(iv) We need a schematic picture to describe the relative amplitudes of vecros of different sizes in our wavefunctional. Recall that in a 1-dimensional potential well, the virtual fluctuation is the part of the wavefunction that is `under the barrier'. In fig.\ref{f4}(a) we draw such a potential well and let the part of the wavefunction under the barrier denote the  `vecro distribution function'. As mentioned above, at this stage the shape of this  vecro distribution function is fixed to a unique shape that  minimizes the energy of the state, so this part of the wavefunctional does not describe any independent excitations.

\begin{figure}
\begin{center}
\includegraphics[scale=.55]{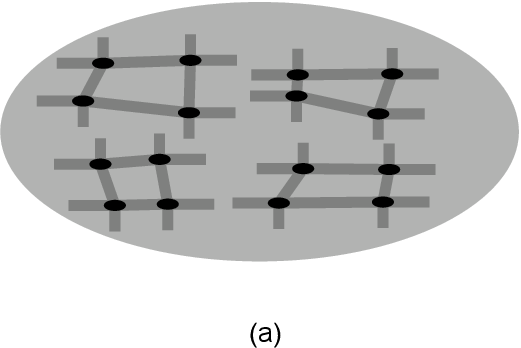}\hskip5em\includegraphics[scale=.55]{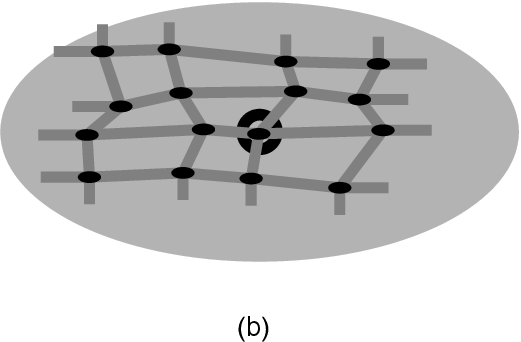}
\end{center}
\caption{(a) Vecro fluctuations in the vacuum. (b) Under the attraction of a central mass (the black annulus), the vecro distribution shifts towards larger vecros.}
\label{f5}
\end{figure}

(v) In fig.\ref{f5} we describe   what happens when instead of the vacuum, we have a dense mass like a star in our spacetime. The vecro configurations of the vacuum in fig.\ref{f5}(a) get attracted and squeezed if we place a heavy object at the center as in fig.\ref{f5}(b).  This squeezing causes the smaller vecros to join up into larger ones. We depict this change schematically in fig.\ref{f4}(b): under the attraction of a heavy central mass, the vecro distribution function shifts towards larger vecros. This shift towards larger vecros continues till the mass of the star reaches the black hole threshold.

(vi) What happens if we add further energy to the star so that it crosses the black hole threshold? The gravitational pull of the star has already caused the vecros to join up till they are almost all of radius $r\approx 2GM$. As we cross the black hole threshold, the wavefunctional over the vecro configuration ceases to be `under the barrier', and becomes an oscillatory function (fig.\ref{f4}(c)). This means that the relative amplitudes of different vecro configurations is not fixed; instead we can have many different wavefunctions over the space of vecro configurations. If the wavefunction is taken to peak around a given configuration, then we get the corresponding fuzzball of fig.\ref{f2}(c). The fuzzball is not a virtual fluctuation: once the wavefunction becomes oscillatory we have real (i.e. on-shell) configurations. Since the space of fuzzball configurations is large, we now get  a large number of allowed states, corresponding to the Bekenstein entropy (\ref{three}). 

(vii) Finally, we should ask what triggers the transition from virtual excitations (vecros) to on-shell fuzzballs exactly as we cross the black hole threshold. The vecro excitations relevant to our dynamics are extended structures that are at rest in the Schwarzschild frame. The gravitational pull in the Schwarzschild frame diverges as $r\r 2GM^+$, so the compression effect on vecros also diverges as we approach the horizon radius. Correspondingly, the distribution of vacuum fluctuations is not smooth in the limit $r\r 2GM^+$: the amplitude for a vecro fluctuation to exist diverges as the size of the vecro is taken towards  $r=2GM^+$. Because of this,  in contrast to the situation in (iii) above, the spectrum of fluctuations is not locally translationally invariant near $r=2GM$. Thus the local vacuum is not the Unruh vacuum (which {\it is} translationally invariant through $r=2GM$) but a state closer to the Boulware vacuum. The Boulware vacuum has a negative Casimir energy, and this balances the positive energy of the fuzzballs which emerge from the vecro fluctuations as in (vi). Note that an initial Unruh horizon can form during rapid gravitational collapse, but the vecros of size $r\approx 2GM$ are static in the Schwarzschild frame and would therefore feel an Unruh acceleration temperature ${1\over 2\pi s}$ at a distance $s$ outside the horizon. As $s\r 0$, this is a diverging excitation of the vecros, and alters their distribution to yield on-shell the fuzzballs mentioned above.


\m

{\bf The picture of infall}

\m

The vecro picture tells us that the traditional black hole geometry is not an allowed solution at mass $M$; the stretching of vecros in such a geometry will always increase its energy above $M$. What we have instead at mass $M$ is a set of $Exp[S_{bek}(M)]$ fuzzball microstates with no horizon. An infalling shell of mass $M$ slows in its evolution as it reaches the horizon radius, since if it continued its collapse through the horizon then it would reach a situation where the energy in the geometry exceeded $M$. A process of tunneling starts into fuzzball states of mass $M$, and the shell turns into a fuzzball which will then radiate from its surface like a normal body.

But one can ask still ask the following questions. Is there any significance to the part of the classical black hole geometry inside the horizon? Can we change to Kruskal coordinates and follow the particles of the shell in their journey to the central singularity? In particular we can ask about the experience of an observer who was sitting at $r=0$ before the shell was sent in. Let the shell move radially inwards at the speed of light. Causality says that the observer at the origin does not feel any change to his evolution until the shell reaches $r=0$. Thus it would appear that there must be a meaning to the region $0<r<2GM$, at least for times of the order of the crossing time. What is the significance of this interior region, and how does its existence fit in with the notion that
the hole is described by fuzzballs?

To understand what happens to such an observer at $r=0$, first recall the well-known phenomenon of alpha-decay. The alpha particle is trapped in a potential well in the nucleus, where it oscillates back and forth with some frequency; we can consider these oscillations as marking the ticking of a `clock'. Let us call this region of the potential as region A. There is a potential barrier trapping the alpha particle, but at some value of the radius $r_0$ the particle emerges from under the barrier and escapes to infinity. Let us call the region $r>r_0$ as region B. 

We start with the alpha particle in region A. Thus the probability $P_A$ for the particle to be in A starts at the value  $P_A=1$, while $P_B=0$. As time passes, $P_A$ decreases towards zero and $P_B$ increases towards unity. Let us now ask: as $P_A$ decreases, what does the clock `feel'? That is, does its period change, indicating damage to the clock? The answer is of course `no', since the oscillations of the alpha particle in the well maintain their period; all that happens is that the probability of the clock existing itself goes to zero as the particle tunnels out. Thus we have a Schrodinger Cat type of situation: the alpha particle is in a superposition of two states (the states in region A and region B). We can in fact call this a Cheshire Cat situation: the alpha particle slowly vanishes from region A, and reappears in region B.

Moving closer to gravity, consider  Minkowski space with an extra compact circle $S^1$. If there are no fermions in the theory, this metric is unstable to tunneling into a `bubble of nothing' \cite{witten}. Let us place a graviton at $r=0$; this is the analogue of the `observer at the origin' and the analogue of the clock in the alpha decay model. What will this graviton `feel' as time evolves? 

We see that the answer is exactly similar to the alpha decay case. To leading order, the overall wavefunctional of the full quantum gravity problem splits into parts. One part with probability $P_A(t)$ describes the graviton evolving normally in flat space in  the region around $r=0$. The other part with probability $P_B(t)=1-P_A(t)$ describes a situation with no spacetime around $r=0$, and an expanding bubble with some radius $R_b$; the bubble metric carries small deformations corresponding to the details of the graviton wavefunction that we started with. Thus we again have a Schrodinger Cat/ Cheshire Cat situation: the part of the wavefunction described by $P_A$ evolves (to leading order) just as if there was no tunneling. This part of the graviton wavefunction does not  get distorted in time due to the bubble nucleation; instead it is the {\it probability} $P_A$ of the graviton being in that state goes smoothly to zero.

A similar situation holds for the black hole. Let there be a graviton at the origin $r=0$, and let a shell of mass $M$ come in at the speed of light from infinity. Consider the situation when the  shell has a radius $R=2GM+\Delta$. The wavefunctional describing the shell has a nonzero overlap with fuzzball microstates of radius $R=2GM+\Delta$. But these fuzzball microstates have a mass 
\be
M_{fuzz}={R\over 2G}=M+{\Delta\over 2G}>M
\label{wone}
\ee
Thus the full quantum gravity wavefunctional is mostly concentrated around the shell state (which has a graviton inside at $r=0$). There is a small penetration `under the barrier' into fuzzball states of mass $M_{fuzz}$; the details of the graviton at $r=0$ is encoded in the detailed shape of this wavefunctional. 

As the shell approaches $R=2GM$, we find that $M_{fuzz}-M$ becomes smaller, and thus the part of the wavefunctional under the barrier becomes larger. At $R\approx 2GM+\l_p$, we have $M_{fuzz}=M$, and the shell can tunnel into the large family of 
$Exp[S_{bek}(M)]$ fuzzball microstates of mass $M$ \cite{tunnel} . At this stage we again have the Schrodinger Cat/Cheshire Cat picture. There is a probability $P_A(t)$ that we have the shell with the graviton inside, and a probability $P_B(t)=1-P_A(t)$ that there is a fuzzball state (i.e., no shell, no interior of a shell, and the details of the graviton encoded in the fuzzball wavefunctional). 

With all this, we can address the puzzling aspects of the classical Penrose diagram. 
The Schwarzschild coordinates end outside the horizon, and cover this outside region for all times $-\infty<t<\infty$. In AdS/CFT duality, the CFT has just one time coordinate, which is  analogous to the Schwarzschild time. This time captures the tunneling into fuzzballs, and so is like our system B in the above examples. But what is the significance of the Kruskal time which captures the inside of  the collapsing shell? We see now that this time describes the evolution of system A, which is the part of the wavefunctional that has not yet tunneled through to fuzzballs. Thus in the fuzzball paradigm the long standing question `do we have both the inside and the outside of the hole?' is resolved by the Schrodinger Cat/Cheshire Cat phenomenon, where we postulate no new physics, but find that the entropy enhanced tunneling into fuzzballs has created two branches of the wavefunctional for our macroscopic system. We need different time coordinates to describe these two branches.

Let us recall what was needed to get the above picture of the black hole interior. First, we need to have fuzzball states (i.e., microstates without horizons) to tunnel to. Second, the vacuum must have a vecro structure of fluctuations, whose deformations in the traditional black hole geometry create a state with energy $E>M$; this forces a tunneling transition from the collapsing shell to the fuzzball states. Third, we need the large Bekenstein entropy $S_{bek}$ to yield an entropy-enhancement of the tunneling so that a macroscopic system like the collapsing shell can transition to a Schrodinger Cat type state with two branches to its wavefunction. The above picture can be shown to get direct support by looking at evolution in the dual CFT; the details of this map will be presented elsewhere.

{\bf Summary:}

\m

Electrons and positrons exist as real particles, so the vacuum must contain corresponding {\it virtual} fluctuations. Recognizing this fact leads to the process of Hawking radiation. This process allows consistency with thermodynamics, but leads to the trap of the information paradox. We escape the trap if we recognize that black holes have a large number of microstates, and thus the vacuum must have corresponding virtual fluctuations. In string theory many of these microstates have been explicitly constructed and are found to have an extended structure with no horizon (fig.\ref{f2}(c)). The corresponding virtual fluctuations, called vecros, give the gravitational vacuum a secret structure where it has correlations at all length scales. In this article we have given an explicit lattice model of these correlations, where the gravitational attraction of a star leads to a squeezing of these vecros; this squeezing  alters the vecro distributions and converts them to on-shell fuzzball states at the black hole threshold. (In \cite{eco} it was shown that such fuzzballs would naturally acquire the Hawking temperature $T_H=1/8\pi GM$.)
Since fuzzballs have no horizon and radiate from their  surface like a normal body, we resolve the information paradox. It is crucial that vecro fluctuations are extended in size, so they are able to `feel around' a region and respond to the formation of a closed trapped surface. By contrast, the traditional picture of the vacuum depicted in fig.\ref{f1} sees nothing special upon horizon formation, and so cannot escape the trap of the paradox. In the fuzzball paradigm we also find a natural explanation for the Kruskal time inside the hole as describing one part of the overall wavefunctional, while the Schwarzschild time captures the other part.

\newpage

 \section*{Acknowledgements}

This work is supported in part by DOE grant DE-SC0011726. I would like to thank Robert Brandenberger, Emil Martinec and Madhur Mehta for several helpful discussions.


\begin{thebibliography}{99}

  
 


 \bibitem{hawking}
  S.~W.~Hawking,
  Commun.\ Math.\ Phys.\  {\bf 43}, 199 (1975)
  [Erratum-ibid.\  {\bf 46}, 206 (1976)];
  S.~W.~Hawking,
  Phys.\ Rev.\  D {\bf 14}, 2460 (1976).
  
\bibitem{cern}
  S.~D.~Mathur,
  Class.\ Quant.\ Grav.\  {\bf 26}, 224001 (2009)
  [arXiv:0909.1038 [hep-th]].
 
  
\bibitem{bek}
J.~D.~Bekenstein,
Phys.\ Rev.\ D {\bf 7}, 2333 (1973).
%
  
 
  
   \bibitem{fuzzballs}
O.~Lunin and S.~D.~Mathur,
  Nucl.\ Phys.\  B {\bf 623}, 342 (2002)
  [arXiv:hep-th/0109154];
O.~Lunin, J.~M.~Maldacena and L.~Maoz,
[arXiv:hep-th/0212210 [hep-th]];
I.~Kanitscheider, K.~Skenderis and M.~Taylor,
  arXiv:0704.0690 [hep-th];
 S.~D.~Mathur,
  Fortsch.\ Phys.\  {\bf 53}, 793 (2005)
  [arXiv:hep-th/0502050];\\
 I.~Bena and N.~P.~Warner,
  Lect.\ Notes Phys.\  {\bf 755}, 1 (2008)
  [arXiv:hep-th/0701216];
  B.~D.~Chowdhury and A.~Virmani,
  ``Modave Lectures on Fuzzballs and Emission from the D1-D5 System,''
  arXiv:1001.1444 [hep-th];
I.~Bena, S.~Giusto, R.~Russo, M.~Shigemori and N.~P.~Warner,
JHEP \textbf{05}, 110 (2015)
[arXiv:1503.01463 [hep-th]];
I.~Bena, S.~Giusto, E.~J.~Martinec, R.~Russo, M.~Shigemori, D.~Turton and N.~P.~Warner,
Phys. Rev. Lett. \textbf{117}, no.20, 201601 (2016)
[arXiv:1607.03908 [hep-th]].
S.~D.~Mathur,
doi:10.1142/S0218271820300098
[arXiv:2001.11057 [hep-th]].




  
\bibitem{ghm}
B.~Guo, S.~Hampton and S.~D.~Mathur,
JHEP \textbf{07}, 162 (2018)
[arXiv:1711.01617 [hep-th]].

\bibitem{gibbonswarner}
G.~W.~Gibbons and N.~P.~Warner,
Class. Quant. Grav. \textbf{31}, 025016 (2014)
doi:10.1088/0264-9381/31/2/025016
[arXiv:1305.0957 [hep-th]].

\bibitem{prevent}
S.~D.~Mathur,
Int. J. Mod. Phys. D \textbf{25}, no.12, 1644018 (2016)
doi:10.1142/S0218271816440181
[arXiv:1609.05222 [hep-th]].

\bibitem{tunnel}
  S.~D.~Mathur,
  arXiv:0805.3716 [hep-th];
  S.~D.~Mathur,
  Int.\ J.\ Mod.\ Phys.\  D {\bf 18}, 2215 (2009)
  [arXiv:0905.4483 [hep-th]];
  P.~Kraus and S.~D.~Mathur,
  Int.\ J.\ Mod.\ Phys.\ D {\bf 24}, no. 12, 1543003 (2015)
  doi:10.1142/S0218271815430038
  [arXiv:1505.05078 [hep-th]];
  I.~Bena, D.~R.~Mayerson, A.~Puhm and B.~Vercnocke,
  JHEP {\bf 1607}, 031 (2016)
  doi:10.1007/JHEP07(2016)031
  [arXiv:1512.05376 [hep-th]].

\bibitem{vecro}
S.~D.~Mathur,
doi:10.1142/S0218271820300098
[arXiv:2001.11057 [hep-th]].

\bibitem{elastic}
S.~D.~Mathur,
Int. J. Mod. Phys. D \textbf{30}, no.14, 2141001 (2021)
doi:10.1142/S0218271821410017
[arXiv:2105.06963 [hep-th]].

\bibitem{bran}
R.~Brandenberger and G.~A.~Mitchell,
Eur. Phys. J. C \textbf{83}, no.4, 308 (2023)
doi:10.1140/epjc/s10052-023-11501-2
[arXiv:2302.12924 [hep-th]].
 


\bibitem{eco}
S.~D.~Mathur and M.~Mehta,
Int. J. Mod. Phys. D \textbf{32}, no.14, 2341003 (2023)
doi:10.1142/S0218271823410031
[arXiv:2305.12003 [hep-th]];
S.~D.~Mathur and M.~Mehta,
[arXiv:2402.13166 [hep-th]].

\bibitem{witten}
E.~Witten,
Nucl. Phys. B \textbf{195}, 481-492 (1982)
doi:10.1016/0550-3213(82)90007-4





\end{thebibliography}
\end{document}